\documentclass[acmsmall,nonacm]{acmart}

\AtBeginDocument{%
  \providecommand\BibTeX{{%
    \normalfont B\kern-0.5em{\scshape i\kern-0.25em b}\kern-0.8em\TeX}}}

\usepackage{amsmath,amsfonts}
\usepackage{algorithm}
\usepackage{array}
\usepackage{textcomp}
\usepackage{url}
\usepackage{graphicx}

\usepackage{caption}
\usepackage{subcaption}
\usepackage{listings}
\usepackage{multirow}
\usepackage{multicol}
\usepackage{algpseudocode}
\usepackage{xcolor}
\usepackage{wasysym}
\graphicspath{{imgs/}}

\definecolor{lightcyan}{RGB}{163,220,255}
\definecolor{darkgreen}{RGB}{50,200,50}

\newcommand{\hl}[1]{{\setlength{\fboxsep}{0pt}\colorbox{yellow}{#1}}}
\newcommand{\hlcyan}[1]{{\setlength{\fboxsep}{0pt}\colorbox{lightcyan}{#1}}}
\newcommand{\ftyes}[0]{\textcolor{darkgreen}{\CIRCLE}}
\newcommand{\ftno}[0]{\textcolor{red}{\Circle}}
\newcommand{\ftmay}[0]{\textcolor{orange}{\LEFTcircle}}
\newcommand{\fp}[1]{\tilde{#1}}

\newcommand{\figref}[1]{Fig.~\ref{fig:#1}}
\newcommand{\tabref}[1]{Table~\ref{tab:#1}}
\newcommand{\secref}[1]{\S\ref{sec:#1}}

\lstdefinelanguage{ST}
{
	morekeywords={
	case,of,if,then,end_if,end_case,super,function_block,extends,var,
	constant, byte,,end_var,var_input, real,bool,var_output,
	dint,udint,word,dword,array, of,uint,not,adr, program, for, end_for, while, do, end_while, repeat, end_repeat, until, to, by, else, elsif, var_in_out
	},
	otherkeywords={
		:, :=, <>,;,\,.,\[,\],\^,1,2,3,4,5,6,7,8,9,0,TRUE, FALSE, \{attribute,  \'hide\'\}
	},
	keywords=[1]{
		case,of,if,then,end_if,end_case,super,function_block,end_function_block,extends,var,
		constant, byte,,end_var,var_input, real,bool,var_output,
		dint,udint,word,dword,array, of,uint,not,adr, :, :=, <>,;,\,.,\[,\],\^,program, for, end_for, while, do, end_while, repeat, end_repeat, until, to, by, else, elsif, var_in_out
	},
	keywordstyle=[1]\color{blue},
	keywords=[2]{
		1,2,3,4,5,6,7,8,9,0, TRUE, FALSE
	},
	keywordstyle=[2]\color{blue},
	keywords=[3]{
		\{attribute,  \'hide\'\}
	},
	keywordstyle=[3]\color{blue},
	sensitive=false,
	morecomment=[l]{//}, 
	morecomment=[s]{(*}{*)},
	morestring=[b]{"},
	morestring=[b]{'}
}

\begin{document}

\title{REMaQE: Reverse Engineering Math Equations from Executables}


\author{Meet Udeshi}
\email{m.udeshi@nyu.edu}
\orcid{0000-0001-7297-0880}
\affiliation{%
  \institution{ECE, NYU Tandon School of Engineering}
  \city{Brooklyn}
  \state{New York}
  \country{USA}
}

\author{Prashanth Krishnamurthy}
\email{prashanth.krishnamurthy@nyu.edu}
\affiliation{%
  \institution{ECE, NYU Tandon School of Engineering}
  \city{Brooklyn}
  \state{New York}
  \country{USA}
}

\author{Hammond Pearce}
\email{hammond.pearce@unsw.edu.au}
\affiliation{%
  \institution{UNSW School of Computer Science and Engineering}
  \city{Sydney}
  \country{Australia}
}

\author{Ramesh Karri}
\email{rkarri@nyu.edu}
\affiliation{%
  \institution{ECE, NYU Tandon School of Engineering}
  \city{Brooklyn}
  \state{New York}
  \country{USA}
}

\author{Farshad Khorrami}
\email{khorrami@nyu.edu}
\affiliation{%
  \institution{ECE, NYU Tandon School of Engineering}
  \city{Brooklyn}
  \state{New York}
  \country{USA}
}



\renewcommand{\shortauthors}{Udeshi et al.}

\begin{abstract}
Cybersecurity attacks on embedded devices for industrial control systems and cyber-physical systems may cause catastrophic physical damage as well as economic loss.
This could be achieved by infecting device binaries with malware that modifies the physical characteristics of the system operation. 
Mitigating such attacks benefits from reverse engineering tools that recover sufficient semantic knowledge in terms of mathematical equations of the implemented algorithm.
Conventional reverse engineering tools can decompile binaries to low-level code, but offer little semantic insight.
This paper proposes the REMaQE automated framework for reverse engineering of math equations from binary executables. 
Improving over state-of-the-art, REMaQE handles equation parameters accessed via registers, the stack, global memory, or pointers,
and can reverse engineer object-oriented implementations such as C++ classes.
Using REMaQE, we discovered a bug in the Linux kernel thermal monitoring tool ``tmon''.
To evaluate REMaQE, we generate a dataset of 25,096 binaries with math equations implemented in C and Simulink.
REMaQE successfully recovers a semantically matching equation for all 25,096 binaries.
REMaQE executes in 0.48 seconds on average and in up to 2 seconds for complex equations.
Real-time execution enables integration in an interactive math-oriented reverse engineering workflow.

\end{abstract}


\maketitle

\section{Introduction} \label{sec:introduction}


Embedded systems in industrial control systems (ICS) and cyber-physical systems (CPS) are vulnerable to process-aware cyber-attacks that compromise the physical processes under control~\cite{Khorrami16ProcessAware,Konstantinou15CPSSec, Sun2020SoK}.
Such attacks can breach safety-critical requirements, causing real-world destruction and harm.
For example, Stuxnet injected malware in the programmable logic controllers (PLC) of a nuclear facility to adversely modify the centrifuge motor frequencies~\cite{Langner11Stuxnet}. 
The BlackEnergy attack against Ukraine's power grid opened several breakers at once to cause a power outage \cite{Lee16BlackEnergy}.
Covert attacks are also possible: Krishnamurthy et al.~\cite{Krishnamurthy18CPSCovert} develop a stealthy communication channel that uses analog emissions from a CPS. This method maintains the closed-loop process characteristics by factoring in controller dynamics during attack design.
Civilian unmanned aerial vehicles (UAVs, drones) are being targetted to inject malicious software trojans in the controller and allow attackers to gain control of the drone \cite{Altawy2016DroneSurvey}.
Sun~et~al.~\cite{Sun2020SoK} present a survey of attacks on ICS, highlighting more examples of control logic modification attacks that impact PLCs.

One can analyze the cybersecurity of embedded systems by leveraging semantic understanding of the physical process features, such as mathematical models, control algorithms, and dynamic behavior of the process.
For example, Yang~et~al.~\cite{Yang22ControlInvariant} derive control invariants of physical processes using PLC runtime logs and detect changes in these invariants to signal an intrusion.
Badenhop~et~al.~\cite{Badenhop2018ZWave} reverse engineer the proprietary Z-Wave transceiver using static and dynamic analysis to verify communication security properties.
The Trusted Safety Verifier in \cite{McLaughlin14TSV} is implemented as a bump-in-the-wire and verifies safety-critical code for PLCs by recovering a semantic graph of the program and asserting safety-critical properties using model checking.
Bourbouh~et~al.~\cite{Bourbouh2021Lustre} implement a reverse compilation framework from Lustre (low-level synchronous data-flow language) to Simulink (high-level modelling framework) to add semantic context to the Simulink models.

On the offensive side, McLaughlin~\cite{McLaughlin11DynamicMalware} shows that recovering boolean expressions of the PLC control code can automatically generate dynamic malware payloads.
The CLIK attack on PLCs \cite{Kalle19CLIK} decompiles the control logic to high-level automation languages, making malicious code changes to disrupt the physical process.
The Laddis decompiler~\cite{Senthivel2018Laddis} enables the attacker to mount a denial of engineering operations (DEO) attack on a PLC by modifying the ladder logic.

While there are many methods to reverse engineer embedded systems and extract semantic knowledge for cybersecurity, this paper focuses on the analysis and reverse engineering of binary executables of the embedded devices (e.g., PLCs) which interact with physical sensors and actuators.
Reverse engineering the mathematical models and control algorithms implemented in the binaries
can reveal the semantic knowledge necessary to understand these embedded systems.
This semantic knowledge is useful for applications such as
(a) analyzing malicious changes injected in the binary by malware, 
(b) recovering details of legacy systems without source code,
(c) examining adversarial systems, such as UAVs or drones used for reconnaissance,
(d) identifying vulnerabilities in implemented systems for patching or exploitation,
and (e) debugging during implementation of mathematical algorithms.

While binary compilation works well in the forward direction (math equation $\rightarrow$ binary executable), reversing this process is  difficult. Math equations compiled into binaries are deployed on a diverse range of embedded hardware platforms and target a variety of processor architectures.
The implementation of math equations involves platform-specific details which offer no semantic information. 
Disassembly and decompilation tools like IDA Pro \cite{idapro} and Ghidra \cite{ghidra} are  useful for analysis of binaries, but they do not recover semantic information. 
Symbolic execution is useful to gain semantic insight into a binary with dynamic analysis~\cite{Angr2016}, but the recovered semantic information is not presented as human-friendly math equations.
There is thus a need for a framework that can automatically reverse engineer math equations from their binaries and present them in a human-friendly form. To this end, we propose REMaQE.

\subsection{Contributions}

The REMaQE framework leverages symbolic execution along with automatic parameter analysis and algebraic simplification methods to automate the reverse engineering of mathematical equations from binary executables.
REMaQE employs automatic parameter analysis of functions to identify important metadata regarding the function arguments stored in register, stack, global memory, or accessed via pointer. 
According to the best of our knowledge, REMaQE is the \emph{first} work able to reverse engineer math equations from a variety of function implementations, including object-oriented C++ code. 
Simplification of math equations in REMaQE is performed via math-aware algebraic methods.
This overcomes limitations of other approaches such as machine learning methods for equation simplification  and enables REMaQE to simplify complex conditional equations. 
Unlike state-of-the-art methods that look for patterns of well-known algorithms,
REMaQE refines the extracted semantic information and displays it as math equations, enabling general-purpose applications on real-world use cases.
Parameter analysis and algebraic simplification sets REMaQE apart from prior semantic reverse engineering approaches. 
\secref{related_work} discusses the strengths of REMaQE over existing works.
The contributions of this paper are as follows:
\begin{enumerate}
\item The REMaQE framework for reverse engineering mathematical equations from binary executables, which offers two major improvements over existing approaches:
    \begin{enumerate}
    \item Automatic parameter analysis to recognize input, output, and constant parameters in an implemented equation. This enables reverse engineering of object-oriented implementations, such as C++ classes and struct pointer based C functions
    \item Algebraic simplification to transform extracted symbolic expressions into easily understandable math equations.
    This handles much more complex expressions compared to existing approaches that use machine learning methods
    \end{enumerate}
\item A dataset of 25,096 compiled binary executables and their corresponding 3,137 math equations with their implementation as C code and Simulink models. This is suitable for evaluating reverse-engineering tools and is made available at~\cite{DatasetDataPort}
\end{enumerate}


The paper is organized as follows:
\secref{motivation} offers a motivating example of the Linux ``tmon'' controller bug found by REMaQE,
\secref{related_work} reviews related work,
\secref{implementation} details REMaQE's implementation,
\secref{evaluation} explains the evaluation procedure and dataset generation methodology,
\secref{results} reports results,
\secref{ardupilot} and \secref{openplc} showcase two case studies on reverse engineering the ArduPilot~\cite{ArduPilot} auto-pilot C++ firmware and an OpenPLC~\cite{OpenPLC} PID controller using REMaQE,
and \secref{conclusion} concludes and explores future work.

\section{Motivation - Linux Kernel PID Controller} \label{sec:motivation}

\begin{figure}[htpb]
\centering
\begin{subfigure}[t]{0.45\linewidth}
    \begin{lstlisting}[language=C]
void controller_handler(const double xk, double *yk) {
  double ek;
  double p_term, i_term, d_term;
  ek = p_param.t_target - xk; /* error */
  if (ek >= 3.0) {
    syslog(LOG_DEBUG, "PID: %3.1f Below set point %3.1f, stop\n", xk, p_param.t_target);
    controller_reset();
    *yk = 0.0;
    return;
  }
  /* compute intermediate PID terms */
  p_term = -p_param.kp * (xk - xk_1);
  i_term = (*@\hlcyan{p\_param.kp}@*) * p_param.ki * p_param.ts * ek;
  d_term = -(*@\hlcyan{p\_param.kp}@*) * p_param.kd * (xk - 2 * xk_1 + xk_2) / p_param.ts;
  /* compute output */
  *yk += p_term + i_term + d_term;
  /* update sample data */
  (*@\hl{xk\_1 = xk;} \label{ln:tmon_bug1}@*)
  (*@\hl{xk\_2 = xk\_1;} \label{ln:tmon_bug2}@*)
  /* clamp output adjustment range */
  if (*yk < -LIMIT_HIGH) (*@ \label{ln:tmon_clamp_start} @*)
    *yk = -LIMIT_HIGH;
  else if (*yk > -LIMIT_LOW)
    *yk = -LIMIT_LOW; (*@ \label{ln:tmon_clamp_end} @*)
  p_param.y_k = *yk;
  set_ctrl_state(lround(fabs(p_param.y_k)));
}
    \end{lstlisting}
    \caption{C source of \texttt{controller\_handler} in ``tmon''.}
    \label{fig:tmon_pid_c}
\end{subfigure}
\hfill
\begin{subfigure}[t]{0.48\linewidth}
    \begin{lstlisting}[language=C]
void FUN_00015194(double *param_1,undefined4 param_2) {
  double in_d0, dVar1, dVar2;
  if (3.0 <= _DAT_00018448 - in_d0) {
    syslog(7,"PID: %3.1f Below set point %3.1f, stop\n");
    FUN_00015150();
    *(undefined4 *)param_1 = 0;
    *(undefined4 *)((int)param_1 + 4) = 0;
    return;
  }
  dVar2 = (((in_d0 - (_DAT_00018458 + _DAT_00018458)) + _DAT_00018460) * -(_DAT_00018420 * _DAT_00018430)) / DAT_00018438 + -(_DAT_00018420 * (in_d0 - _DAT_00018458)) + _DAT_00018420 * _DAT_00018428 * DAT_00018438 * (_DAT_00018448 - in_d0) + *param_1;
  dVar1 = -95.0;
  *param_1 = dVar2;
  _DAT_00018458 = in_d0;
  _DAT_00018460 = in_d0;
  if (-95.0 <= dVar2) {
    dVar1 = -2.0;
    if (dVar2 < -2.0) goto LAB_00015270;
  }
  *param_1 = dVar1;
LAB_00015270:
  _DAT_00018450 = *param_1;
  lround((double)CONCAT44(param_2,param_1));
  FUN_00014ddc();
  return;
}
    \end{lstlisting}
    \caption{\texttt{controller\_handler} decompiled with Ghidra.}
    \label{fig:tmon_pid_ghidra}
\end{subfigure}

\begin{subfigure}[t]{0.95\linewidth}
    \begin{lstlisting}[numbers=none]
<FP64 fpToFP((if reg_r0_4_32{UNINITIALIZED} == 0xffffffff && reg_r0_4_32{UNINITIALIZED} + 0x4 == 0x3 && (1 & ~(<...>[0:0] & 1 ^ <...>[0:0] & 1)) == 1 then reg_d0_1_64{UNINITIALIZED} else (if ((LShR(<...>, <...>)[0:0] | <...>[0:0] ^ <...>[0:0]) & 1) == 1 && reg_r0_4_32{UNINITIALIZED} == 0xfffffff0 && ((LShR(<...>, <...>)[0:0] ^ 1) & 1) != 0 && (1 & ~(<...> & <...> ^ <...> & <...>)) != 1 then fpToIEEEBV(fpAbs(fpAdd(RM.RM_NearestTiesEven, fpAdd(RM.RM_NearestTiesEven, fpDiv(RM.RM_NearestTiesEven, fpMul(RM.RM_NearestTiesEven, <...>, <...>), FPV(0.0, DOUBLE)), fpAdd(RM.RM_NearestTiesEven, fpNeg(<...>), fpMul(RM.RM_NearestTiesEven, <...>, <...>))), fpToFP(mem_fffffff0_5_64{UNINITIALIZED}, DOUBLE)))) else 0x0)), DOUBLE)>
    \end{lstlisting}
    \caption{Symbolic expression generated with Angr.}
    \label{fig:tmon_pid_angr}
\end{subfigure}

\begin{subfigure}[b]{\linewidth}
\centering
\begin{minipage}{0.65\linewidth}
\begin{equation*}
t = x_8 - \frac{\textcolor{cyan}{x_{1}} x_{3}}{x_4} \left(x_{0} - 2 x_{6} + x_{7}\right)
  - \textcolor{cyan}{x_{1}} x_{2} x_{4} \left(x_{0} - x_{5}\right)
  - x_{1} \left(x_{0} - x_{6}\right)
\end{equation*}
\begin{equation*}
y_2 = \begin{cases} t & \text{for}\: x_5 - x_0 \leq k_2 \:\text{and}\: k_{4} \geq t \:\text{and}\: k_{3} \leq t \\k_{3} & \text{for}\: x_5 - x_0 \leq k_2 \:\text{and}\: k_{3} > t \\k_{4} & \text{for}\: x_5 - x_0 \leq k_2 \:\text{and}\: k_{3} \leq t \:\text{and}\: k_{4} < t \\0 & \text{otherwise} \end{cases}
\end{equation*}
\end{minipage}
\begin{minipage}{0.3\linewidth}
\begin{equation*}
y_3 = \begin{cases} x_{0} & \text{for}\: x_5 - x_0 \leq k_2 \\0 & \text{otherwise} \end{cases}
\end{equation*}
\begin{equation*}
y_4 = \begin{cases} x_{0} & \text{for}\: x_5 - x_0 \leq k_2 \\0 & \text{otherwise} \end{cases}
\end{equation*}
\end{minipage}
    \caption{Math equations recovered with REMaQE. The term $t$ has been introduced to display the $y_2$ equation clearly.}
    \label{fig:tmon_pid_remaqe}
\end{subfigure}

\caption{
Reverse engineering the Linux ``tmon'' thermal controller with different tools:
(a) C source code,
(b) decompilation with Ghidra,
(c) symbolic execution with Angr,
(d) math equations recovered with REMaQE.}
\label{fig:tmon_example}
\end{figure}

This section presents the reverse engineering of the Linux kernel thermal monitoring tool ``tmon''%
\footnote{\url{https://github.com/torvalds/linux/blob/v6.3/tools/thermal/tmon/pid.c}} that uses a Proportional-Integral-Derivative (PID) controller.
We demonstrate how the math equations recovered by REMaQE help in uncovering a bug in the controller's implementation.
\figref{tmon_pid_c} shows the source code of the \texttt{controller\_handler} function that implements the PID controller.
The source code is provided only for discussion and is not utilized during reverse engineering.
A pre-compiled binary for the ARM 32-bit Hard-Float (ARM32-HF) target is taken from the package repository of the Alpine Linux distribution%
\footnote{\url{https://dl-cdn.alpinelinux.org/alpine/v3.18/community/armhf/linux-tools-tmon-6.3.12-r0.apk}}, which is popularly used on embedded platforms like Raspberry Pi.
For a comparison, the generated binary is reverse engineered using different tools namely, Ghidra~\cite{ghidra} for decompilation, Angr~\cite{Angr2016} for symbolic execution, and REMaQE to recover math equations.

The function is decompiled using Ghidra to generate the C source shown in \figref{tmon_pid_ghidra}.
The decompiled \texttt{controller\_handler} function is presented in a complicated form by Ghidra - it is cluttered with global variable access, struct pointer dereferencing and \texttt{goto} statements for control flow.
This makes it difficult to understand what is implemented, even with the knowledge that the function contains a PID controller.
Ghidra has generated a faithful representation.
However, these implementation details present in the syntax are obscuring the semantic information of this function.
Thus, significant effort is required to clean up and extract a math equation of the PID controller.

Note that the function name is not recovered as the pre-compiled binary is stripped of symbols.
The user needs to identify the function of interest to analyze by looking at either the disassembly or the decompiled output of Ghidra.
In this manner, REMaQE is designed to be used in conjunction with existing reverse engineering tools.
\figref{tmon_pid_angr} shows the symbolic expression generated by Angr.
Even though the generated expression has recovered sufficient semantic information about the source function, 
the implemented equations are not readily apparent to the human analyst.

\figref{tmon_pid_remaqe} shows the math equations recovered by REMaQE for the various output parameters of the function. 
REMaQE can not recover variable names from the stripped binary; hence it names all inputs, outputs, and constants as $x_n$, $y_n$, and $k_n$ respectively.
For this example, REMaQE has identified the function parameters \texttt{*yk}, \texttt{xk\_1}, \texttt{xk\_2} as inputs $x_8$, $x_6$, $x_7$ and outputs $y_2$, $y_3$, $y_4$ respectively.
These three variables are accessed in different ways (pointer dereference, global access) and they have been used as both inputs and outputs in the function; yet, REMaQE has uniformly presented the output equations clearly. This is discussed in detail in \secref{parameter_analysis}.

From the equations recovered by REMaQE, we see that $y_3$ and $y_4$ will take the same value after the function is run once. When the function is called again, variables \texttt{xk\_1} and \texttt{xk\_2}, now used as inputs $x_6$ and $x_7$, will be the same. In the PID control equation,  the D-term $\frac{x_1 x_3}{x_4}(x_0-2x_6+x_7)$, will simplify to $\frac{x_1 x_3}{x_4}(x_0-x_6)$.
This degrades the three-point approximation of the D-term to a two-point one, impacting the quality of the PID controller.
The reason is clear from the yellow highlighted lines in \figref{tmon_pid_c}: line~\ref{ln:tmon_bug1} wrongly reassigns \texttt{xk\_1 = xk} before line~\ref{ln:tmon_bug2} assigns \texttt{xk\_2 = xk\_1}, which means that both variables take the value \texttt{xk}.
The correct implementation would be to switch the order of assignment and assign \texttt{xk\_2} before reassigning \texttt{xk\_1}.
This can be introduced due to either human error or malware which swaps the order of a few assembly instructions.
The impact on the code and the execution is insignificant, yet the physical characteristics of the controller change noticeably and may have real-world consequences.
We have submitted a patch to fix this bug\footnote{\url{https://patchwork.kernel.org/project/linux-pm/patch/20230822184940.31316-1-mudeshi1209@gmail.com/}}.
This example demonstrates the advantage of reverse engineering using REMaQE by providing a semantically rich understanding of the implemented math equations for binary analysis.

\section{Related Work} \label{sec:related_work}

As discussed in \secref{introduction} and \secref{motivation}, disassembly and decompilation techniques 
are unable to provide relevant semantic information of a program, while REMaQE is able to 
portray the program as semantically equivalent math equations.
The goal of decompilation is to reconstruct the original source code from binary executables as accurately as possible. 
Decompilers intend to recover many implementation details of the program such as variable types, memory layout, control flow, and data flow \cite{cifuentes1994reverse}.
These syntactical details are necessary when the recovered programs need to be represented in high-level source languages like C or Java.
However, due to the noisy and obfuscating nature of compilation, decompilation is an indeterminate process; hence, the recovered code may obscure the original high-level semantic meaning.
The reverse engineering approach of REMaQE aims to refine and discard these implementation details before generating human-friendly equations using conventional math symbols and operations.
REMaQE is not intended to replace decompilers or decompilation techniques. 
Instead, it can be integrated with the user interface of decompilation tools to offer
the semantic information of equations alongside useful information extracted by the decompiler, 
for example, the C code of a function along with the math equation.

Manually recovering math equations from the binaries or obscured decompilation outputs requires subject matter expertise of identifying and clustering code sequences, mapping them to math primitives, and understanding how identified blocks will combine and simplify.
Automated approaches tailored to the embedded systems can use knowledge about the compilation tool chain and build a better representation in domain-specific languages.
For example, executables from control devices can be reverse engineered to automation languages like Ladder Logic and Instruction List~\cite{Senthivel2018Laddis,Kalle19CLIK,Qasim2019,Lv2017}.
These methods offer better abstraction than languages like assembly or C, but they too cannot represent the semantic information. This  necessitates  domain expertise and manual effort to recover math equations.

Programs may be intentionally obfuscated. For example, malware is obfuscated to hide its intent and avoid detection or analysis \cite{Kane11Obfuscation}.
Jha~et~al.~\cite{Jha10Oracle} use oracle-guided program synthesis to develop a semantic understanding to deobfuscate malware.
Malware obfuscation uses bit-manipulation operations and deliberate complex control flows specifically to hinder analysis and prevent reverse engineering.
Deobfuscation is outside the scope of the current paper (see \secref{assumptions}).

Symbolic execution is a technique to gain semantic insight into a binary with dynamic analysis.
Symbolic execution explores all execution paths through the program and captures symbolic expressions~\cite{Angr2016}.
Primary use cases are automatic test generation ~\cite{KLEE08,stephens2016driller},
exploit detection and generation~\cite{Mayhem12,firmalice2015}, and reverse engineering~\cite{David16BINSEC, McLaughlin14TSV}.
The symbolic expressions are ``solved'' using satisfiability modulo theory (SMT) solvers to identify concrete inputs.
Although symbolic expressions are generated to assist SMT solvers, they are not presented as human-friendly  equations.
They contain platform-specific implementation details that obscure the semantic information (see \secref{simplify_eq}).
Simply printing the symbolic expressions in a human-readable format is not sufficient.
REMaQE employs algebraic simplification techniques to clean up the expressions.

Recent approaches to extract high-level semantic information \cite{Sun19MISMO, Kim22Dispatch, keliris2019icsref}
have used domain knowledge and assumptions to match the extracted information to patterns of well-known algorithms.
MISMO~\cite{Sun19MISMO} performs 
semantic matching on expressions extracted using symbolic execution to determine which control algorithm is implemented.
DisPatch~\cite{Kim22Dispatch} targets controllers for robotic aerial vehicles, and identifies customized implementations of the PID controller.

MISMO authors have analyzed the ``tmon'' tool and identified that the implemented PID control equations do not match the expected pattern.
In \figref{tmon_pid_c}, the blue highlighted term \texttt{p\_param.kp} is multiplied for the I and D terms, due to which their pattern match fails.
Using REMaQE, the user can identify this mismatch, as highlighted by the blue terms \textcolor{cyan}{$x_1$} in the equations in \figref{tmon_pid_remaqe}.
However, MISMO \emph{fails} to identify the wrong assignment to \texttt{xk\_1} and \texttt{xk\_2} that causes the bug described in \secref{motivation}.
Prior methods are  domain-specific and can be sensitive to alterations in implementation that breach domain-specific assumptions. Although they extract semantic information similar to REMaQE, implementing and deploying such techniques necessitates combining domain expert knowledge with reverse engineering. These methods may fail on outlier cases.
In contrast, REMaQE provides a versatile approach based on a fundamental set of assumptions that can be readily expanded, enabling its application to various domains with minimal adaptation.


PERFUME~\cite{Weideman21PERFUME} 
overcomes limitations in prior domain-specific approaches by following a generic approach which does not rely on semantic pattern matching. PERFUME is able to present the
math equation pertaining to the implemented algorithm in a human-friendly form.
Among existing works, PERFUME is most similar to REMaQE.
PERFUME uses symbolic execution to extract the semantic information in the form of symbolic expressions, and
trains a machine learning based sequence-to-sequence machine translation model to ``translate'' 
that symbolic expression into a simplified equation. The framework
is offered as a plugin for Ghidra that augments decompiled output with  semantic information and 
integrates into an interactive reverse engineering workflow.

PERFUME, however, cannot analyze functions with parameters accessed through the stack, global memory, or pointers. Hence, PERFUME fails to analyze the ``tmon'' tool in \secref{motivation}, because this function uses a struct, a pointer dereference, and global variables to read inputs and write outputs. In contrast, REMaQE employs a parameter analysis pass collecting metadata about each parameter's storage location.
REMaQE can reverse engineer a wider variety of functions, including those found in object-oriented implementations in real-world programs like ``tmon''.
Additionally, while PERFUME's machine translation model struggles to simplify long and complex symbolic expressions arising from conditional logic, REMaQE's algebraic simplification stage adopts a math-aware algorithmic approach. It handles complex expressions generated from functions with floating-point comparisons and conditional logic. This deterministic simplification scales based on the complexity of the simplified expression, not the input expression.
PERFUME's machine translation model cannot simplify the complex conditional expression (\figref{tmon_pid_angr}) generated for the output clamping logic of the ``tmon'' controller (\figref{tmon_pid_c}, lines~\ref{ln:tmon_clamp_start}--\ref{ln:tmon_clamp_end}), whereas REMaQE produces a compact, human-friendly equation.
REMaQE's parameter analysis and algebraic simplification features set it apart from PERFUME and other symbolic execution approaches.

\begin{table}[tb]
    \caption{Feature comparison of REMaQE with existing approaches for reverse engineering math equations.
    \ftyes=fully supported, \ftmay=supported for some cases, \ftno=not supported.} 
    \label{tab:ft_comp}
    \centering
    \begin{tabular}{|c|c|c|c|c|c|}
        \cline{3-6}
        \multicolumn{2}{c|}{} & DisPatch \cite{Kim22Dispatch} & MISMO \cite{Sun19MISMO} & PERFUME \cite{Weideman21PERFUME} & REMaQE (our) \\
        \hline
        \multicolumn{2}{|c|}{\bf Fixed patterns} &    \ftyes  & \ftyes & \ftyes & \ftyes \\ \hline
        \multicolumn{2}{|c|}{\bf General purpose}   & \ftno   & \ftno  & \ftyes & \ftyes \\ \hline
        \multirow{4}{0.12\linewidth}{\bf Parameter Access}
                                          & register    & \ftmay & \ftmay & \ftyes & \ftyes \\ \cline{2-6}
                                          & stack  & \ftmay & \ftmay & \ftno  & \ftyes \\ \cline{2-6}
                                          & pointer& \ftmay & \ftmay & \ftno  & \ftyes \\ \cline{2-6}
                                          & global & \ftmay & \ftmay & \ftno  & \ftyes \\ 
        \hline
        \multicolumn{2}{|c|}{\bf Conditionals}     & \ftno  & \ftmay & \ftno & \ftyes \\
        \hline
    \end{tabular}
\end{table}

\tabref{ft_comp} compares the features of existing approaches and REMaQE. DisPatch and MISMO require pre-programmed patterns of well-known control algorithms and extract information pertaining to the matched pattern.
They support all kinds of parameter access and conditional statements, but as part of the matched pattern. On the other hand, PERFUME is general-purpose, however it can only handle parameters present in registers, severely restricting its scope. REMaQE is general-purpose, yet handles all kinds of parameter access. It also handles conditional logic in a general context.

\section{Implementing REMaQE} \label{sec:implementation}

\subsection{Background on Symbolic Execution} \label{sec:symbolic_exec}

Symbolic execution involves running a program in an emulated environment using symbols instead of concrete values. This process generates a symbolic expression tree (ET), a type of abstract syntax tree. When conditional branches depend on symbolic data, the execution forks to follow both taken and not taken paths, adding a constraint or predicate to each path based on the branch condition. Constraints accumulate as the execution progresses through each branch. Upon reaching an equivalence point, such as a function return statement, the ETs can be combined into one, utilizing the constraints gathered along each path.
For example, symbolic execution of the binary code of \texttt{controller\_handler} shown in \figref{tmon_pid_c}  forks paths on the branch instruction on line~\ref{ln:tmon_clamp_start}, leading to two execution paths, one which takes the branch and the other which does not.
The paths are merged into one at the function return, and the ET for the return value is simplified by REMaQE into a piecewise form as shown in \figref{tmon_pid_remaqe}.

When reverse engineering math equations, symbolic execution offers two big advantages.
First, operations representing the math equation are naturally captured and tracked. 
Second, the forking on branches ensures that all code is explored during execution, yielding a full picture of the implemented math equation.
Code coverage is important, as failing to explore the entire code base can yield a math equation that is undefined for certain input values.
REMaQE relies on the Angr binary analysis framework~\cite{Angr2016} for dynamic analysis and symbolic execution.


\subsection{Overview of REMaQE}

\begin{figure*}[ht]
    \centering
    \includegraphics[width=\linewidth]{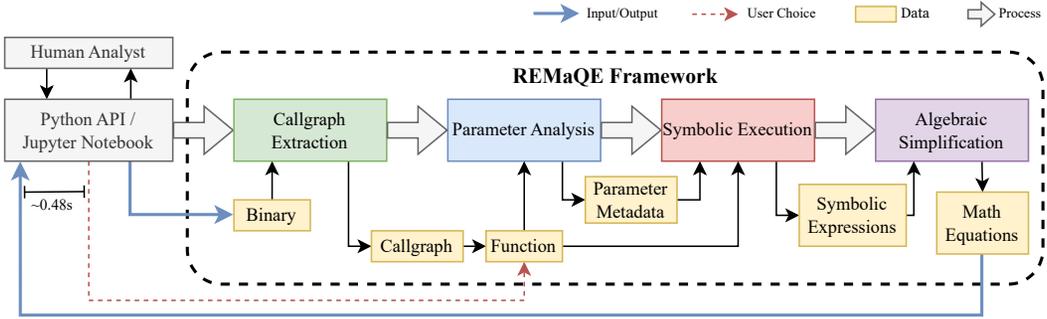}
    \caption{The REMaQE framework. Average execution time of the pipeline is 0.48 seconds, from when the user provides which function to reverse engineer, to when REMaQE returns the math equations.}
    \label{fig:overview}
\end{figure*}

\figref{overview} presents an overview of the REMaQE framework.
REMaQE focuses on reverse engineering math equations of one function at a time.
This allows reverse engineering to proceed in a modular fashion,
similar to the approach a human analyst may follow manually.
REMaQE first extracts a call graph of the functions present in the executable.
The user then selects a function for REMaQE to analyze.
The function runs through the parameter analysis stage
to gather information about input, output and constant parameters that express
the math equation. This information is packaged into the parameter metadata.
The symbolic execution stage then runs the function with initialized inputs and extracts the symbolic ETs for each output.
The output ETs are simplified by the algebraic simplification stage to generate the math equation.

REMaQE's function-by-function approach enables control over whether to represent function calls as-is or substitute them in the recovered equation.
For instance, trigonometric functions might be implemented using approximating polynomials or platform-specific hardware extensions in libraries.
However, such implementation details do not provide valuable information and may even cloud the equation's meaning.
In such cases, it is better to represent these as function calls in the generated equation.
Conversely, in some situations, it could be beneficial to substitute a called function's equation into the caller function's equation.
The \texttt{controller\_reset} function in ``tmon'' is one such example, where it is beneficial to be substituted.
Certain functions which do not impact the output can be ignored, such as logging, printing, and error functions (e.g., \texttt{syslog}).

REMaQE provides an application programming interface (API) in Python,
which allows the user to analyze the executable function by function and to control each stage of REMaQE.
The API can be used in Python scripts or in a Python command line interface like I-Python or Jupyter for interactive reverse engineering.
It can also be integrated as a plugin into  decompilation frameworks with a graphical interface such as Ghidra (to be addressed in future work).

\begin{figure}[htb]
\centering
\includegraphics[width=\linewidth]{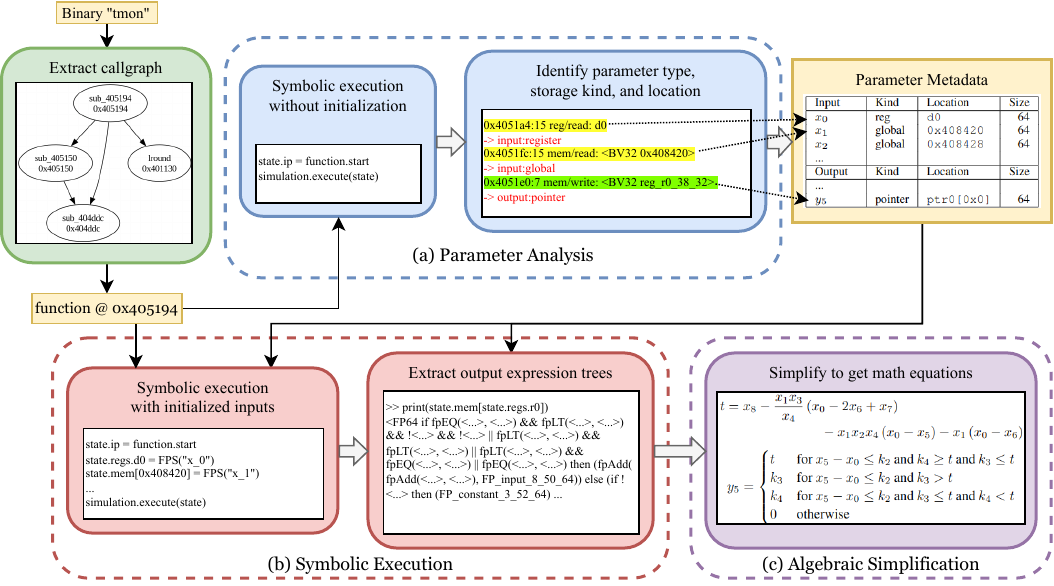}
\caption{REMaQE reverse engineering pipeline. The \texttt{controller\_handler} function from \figref{tmon_example} is used as an example, and intermediate outputs are shown for each stage through the pipeline: 
(a) Parameter analysis automatically identifies the input, output and constant parameters of the function along with their kind and storage location,
(b) Symbolic execution runs the function with properly initialized symbolic inputs and gathers the output symbolic ETs, and
(c) Algebraic simplification converts the output ET to a  human-friendly math equation.}
\label{fig:flowdiagram}
\end{figure}

\figref{flowdiagram} shows the REMaQE pipeline. 
The Linux ``tmon'' example discussed in \secref{motivation} is used to explain the REMaQE pipeline.
The \texttt{controller\_handler} function is selected for analysis.
Snippets of intermediate outputs of each stage are elaborated in 
\secref{parameter_analysis}, \secref{rq_symbolic_exec}, and \secref{simplify_eq}.

\subsection{Parameter Analysis} \label{sec:parameter_analysis}

To invoke symbolic execution of a function, it is important that the execution entry state is initialized with symbolic parameters placed in the expected locations.
To read the output ET, it is necessary to know where the function places the outputs. A math equation may use certain constants, which the tool should distinguish from inputs and extract their hard-coded value. Parameter analysis stage collects the information of the storage location of parameters.

The parameters of a function are defined as the inputs, outputs and constants which are used to express the implemented math equation, and pointers which hold the storage address of the parameters.
Temporary variables which may be used in the function implementation are not considered as parameters. 
In the ``tmon'' example, the \texttt{controller\_handler} function takes a pointer to \texttt{yk} as an argument.
This pointer is not considered an input/output parameter by REMaQE, instead it is separately marked and the dereferenced value is considered as input/output.
Similarly, the global variables \texttt{xk\_1} and \texttt{xk\_2} are considered as output parameters as they are written to by the function, even though they are not part of the function's return value.

The memory layout of data at the pointer address or inside a struct is not important since each separate value, whether it is dereferenced or accessed as a member, is considered a distinct parameter.
Only the offset is necessary.
This distinction helps to abstract different implementation details, allowing for a clean math equation recovery.
The metadata for each parameter contains the following information: (i) Parameter name, (ii) Storage kind, and  (iii) Storage location.
Storage location, depending on kind, is either the architectural name of {\bf register}, {\bf  stack} offset, {\bf  global} address, or {\bf pointer} address and offset. 
The pointer parameters also require a storage location, since a concrete address pointing to memory containing symbolic parameters must be passed as an argument to the function.
Constants may also be present as hard-coded immediate values in the instruction. This is indicated as the \textbf{immediate} kind, with the instruction address as location.

\begin{algorithm}[tb]
\caption{Parameter analysis.}
\label{alg:param_analysis}
\begin{algorithmic}[1]
\Procedure{AnalyzeParams}{$F$: function}
\State $T$: execution trace $\gets$ SymbolicExecute$(F)$
\State \emph{Inputs, Outputs, Constants} $\gets$ empty lists
\For{$A$: action {\bf in} $T$}
    \State $L \gets $ StorageLocation$(A)$
    \State \emph{Reads}, \emph{Writes} $\gets$ GetReadsWrites$(L)$
    \If{\emph{Writes} is empty {\bf and} $L$ is initialized}
        \State $V \gets $ GetValue$(L)$
        \State Append $(L, V)$ to \emph{Constants}
    \ElsIf{first(\emph{Reads}) $<$   first(\emph{Writes})}
        \State Append $L$ to \emph{Inputs}
    \EndIf
    \If{\emph{Writes} is not empty}
        \State Append $L$ to \emph{Outputs}
    \EndIf
\EndFor
\State \Return{\emph{Inputs, Outputs, Constants}}
\EndProcedure
\end{algorithmic}
\end{algorithm}

Algorithm~\ref{alg:param_analysis} describes parameter analysis of a function.
The parameter analysis pass has no knowledge of function parameters, so it does not initialize the symbolic execution state and relies on Angr to fill in uninitialized values when accessed. 
The ``SymbolicExecute'' call performs this uninitialized execution and records an execution trace. The execution trace is parsed and read/write accesses to each  storage location are recorded in a sequential history. ``GetReadsWrites'' gathers this access history for each storage location.
Locations with a first access as read are marked as inputs. These input locations are filtered and labeled as constant if they have no write access and contain an initialized value. Uninitialized locations are filled with a symbolic value by Angr, differentiating inputs from constants since constant parameters are hard-coded in the binary.
Some constants that are initialized at runtime (e.g., C++ class member assigned in the the class constructor) will be treated as inputs because symbolic execution begins at the entry point of the selected function, and initializations outside the function cannot be tracked. This does not affect the recovered equation  except that the constant is treated as input and it's value cannot be extracted.
Locations with at least one write access are marked as output. A single location can be both input and output, as the implementation may reuse storage locations.


\begin{table}[tpb]
\caption{Parameter metadata generated for the ``tmon'' PID controller in \figref{tmon_example}. The input, output, constant and pointer parameters are listed, along with storage kind and location.}
\label{tab:tmon_metadata}

\begin{minipage}{0.45\linewidth}
\centering
\begin{tabular}{|l|l|l|r|}
\hline
\textbf{Input}   & \textbf{Kind}    & \textbf{Location}       &   \textbf{Size}  \\
\hline
$x_0$     & register & \texttt{d0      }       &     64  \\
$x_1$     & global  & \texttt{0x408420}       &     64  \\
$x_2$     & global  & \texttt{0x408428}       &     64  \\
$x_3$     & global  & \texttt{0x408430}       &     64  \\
$x_4$     & global  & \texttt{0x408438}       &     64  \\
$x_5$     & global  & \texttt{0x408448}       &     64  \\
$x_6$     & global  & \texttt{0x408458}       &     64  \\
$x_7$     & global  & \texttt{0x408460}       &     64  \\
$x_8$     & pointer & \texttt{ptr0[0x0]} &     64 \\
\hline
\textbf{Pointer}   & \textbf{Kind}   & \textbf{Location}   &   \textbf{Size} \\
\hline
\texttt{ptr0} & register    & \texttt{r0}         &     32 \\
\hline
\end{tabular}
\end{minipage}
\begin{minipage}{0.54\linewidth}
\centering
\begin{tabular}{|l|l|l|r|r|}
\hline
    \textbf{Output}   & \textbf{Kind}    & \textbf{Location}       &   \textbf{Size} & \\
\hline
$y_0 $     & register     &   \texttt{r0}       &     32 & \\
$y_1 $     & register     &   \texttt{d0}       &     64 & \\
$y_2$     & global  & \texttt{0x408450} &     64 & \\
$y_3$     & global  & \texttt{0x408458} &     64 & \\
$y_4$     & global  & \texttt{0x408460} &     64 & \\
$y_5$     & pointer & \texttt{ptr0[0x0]}&     64 & \\
\hline
 \textbf{Constant}   & \textbf{Kind}   & \textbf{Location}   &   \textbf{Size} &   \textbf{Value} \\
\hline
$k_2$        & global & \texttt{0x405298}   &     64 &       3 \\
$k_3$        & global & \texttt{0x4052a0}   &     64 &     -95 \\
$k_4$        & global & \texttt{0x4052a8}   &     64 &      -2 \\
\hline
\end{tabular}
\end{minipage}

\end{table}

\tabref{tmon_metadata} shows the parameter metadata obtained for the ``tmon'' \texttt{controller\_handler} function.
Input $x_0$ is assigned to the ARM 64-bit floating point register \texttt{d0}.
Inputs $x_1$ to $x_7$ are loaded from global memory. Input $x_8$ is dereferenced from the pointer \texttt{ptr0} with offset 0.
As indicated by the pointer table, \texttt{ptr0} is present in 32-bit integer register \texttt{r0}.
Of the 6 outputs, 5 have the same location as inputs, yet they are distinguished as separate parameters of the equations.
During the subsequent symbolic execution stage, register \texttt{d0} is initialized as the symbolic input $x_0$ at start and the ET for $y_4$ is read from it at the end.
Similarly, the outputs $y_0, y_2, y_3$ are co-located with inputs $x_8, x_6, x_7$ and are handled accordingly.

\subsection{Symbolic Execution} \label{sec:rq_symbolic_exec}

This stage initializes the symbolic state with appropriate symbols for each parameter,
using the metadata generated during parameter analysis. 
The execution runs till the function returns, at which point all execution paths are merged into a final state. The parameter metadata is used to read the output ET from the correct storage locations in the final state. Due to the proper initialization in this stage, the output ET
depends on the appropriate symbols, hence a math equation can
be derived for the output.
Constants are treated as symbols during reverse engineering and are optionally substituted in the final equation, so that the constant values are not changed by any computation during execution or simplification. 
REMaQE recovers an accurate depiction of the constants which are hard-coded in the binary and how they are expressed in the math equation,
including modifications during compilation like floating point approximation.

\subsection{Algebraic Simplification} \label{sec:simplify_eq}


The equations generated by reverse engineering have operations which follows the sequence of computations performed by the function's binary implementation.
This representation may be cluttered with implementation-specific operations that need to be cleaned up and simplified to produced human-friendly equations.
Conditional branches also generated complex expressions with deeply nested \texttt{if-then-else} statements.
Such expressions can result when merging execution paths with constraints as described in \secref{symbolic_exec}.
\figref{tmon_pid_angr} shows the internal representation of the ET obtained after symbolic execution of \texttt{controller\_handler}.
The complex expression is generated because of the conditional statements executed for the clamping operation in  lines~\ref{ln:tmon_clamp_start}--\ref{ln:tmon_clamp_end}.
Even though the clamping requires only two conditions, the generated ET represents the sequence of operations performed during symbolic execution which semantically represent the conditions  in the function.

The sequence of operations is determined by two factors.
First, when the C function is compiled to a target machine code, the conditional branches can be represented using a variety of instruction sequences as decided by the compiler.  
Second, Angr maps every instruction in the binary from the target's machine code to an intermediate representation required by it's symbolic execution engine. 
This further modifies the sequence of operations from what the binary's instructions describe.
The only guarantee is that these modifications and mappings are semantically correct and represent the original equation.
When the condition involves a comparison, a separate instruction performs the comparison and updates bits in a flag register.
Later, conditional instructions check these bits of the flag register.
When these instructions are mapped to the IR, Angr can generate bit-wise operations to perform checks, so these operations end up in the final expression.

REMaQE uses algebraic simplification to clean up implementation-specific operations and represent them as a simplified equation.
The if-then-else operation of Angr requires a condition, a true clause, and a false clause. The condition ET must yield a boolean value.
In math equations, it is logical to presume that the condition consists of one or more comparisons combined with boolean operations.
Algebraic simplification exploits this to streamline the condition ET to simple comparisons before producing the equation. 

Each condition ET in the full expression is simplified by Algorithm~\ref{alg:simplification}.
The ``CanonicalOrder'' function gives an ordered pair of the left-hand and right-hand side of the comparison $C$.
Two booleans, $LT$ and $EQ$, are generated to represent less-than and equal-to comparisons.
They are combined to form the representative boolean expression $B$ by the ``BoolExp'' function, and $C$ is replaced with $B$.
The ordering of $LHS$ and $RHS$, along with representing the comparison using $LT$ and $EQ$, helps generate a canonical form of the comparison.
For example, the condition \texttt{(x < y) \&\& (y >= x)} converts to \texttt{LT\_x\_y \&\& (LT\_x\_y || EQ\_x\_y)}, which simplifies to \texttt{LT\_x\_y} and converts back to \texttt{x < y}. 
The Quine-McCluskey algorithm~\cite{Quine, McCluskey} is used to simplify the boolean expression $ET$ to $S$. 
This algorithm does not scale to expressions with a large number of boolean variables. 
However, based on our assumption that the simplified equations involve only a few comparisons, we typically handle a small number of boolean variables, even though the unsimplified expression may be long and complex with many terms.
The boolean variables in $S$ are replaced back with the equivalent floating-point comparisons.

\begin{algorithm}[htpb]
\caption{Simplify conditional ETs.}
\label{alg:simplification}
\begin{algorithmic}[1]
\Procedure{SimplifyConditional}{$ET$}
\For{{\bf each} comparison $C$ {\bf in} $ET$}
    \State $LHS, RHS \gets$ CanonicalOrder$(C)$
    \State $LT \gets LHS < RHS$
    \State $EQ \gets LHS == RHS$
    \State $BE \gets$ BoolExp$(C, LT, EQ)$
    \State Replace $C$ with $BE$ in $ET$
\EndFor
\State $S \gets$ QuineMcCluskey$(ET)$
\For{{\bf each} boolean $B$ {\bf in} $S$}
    \State $C \gets $ GetComparison$(B)$
    \State Replace $B$ with $C$ in $S$
\EndFor
\State \Return{$S$}
\EndProcedure
\end{algorithmic}
\end{algorithm}

The equations are further reduced using the Sympy symbolic processing engine~\cite{sympy}.
Sympy applies rule-based modifications to simplify or cancel terms in the equation.
The definition of simplification is quite subjective, so equations may have more than one representation which can be deemed as simplified. 
Sympy uses the number of operations as a heuristic to quantify simplification level.
Sympy's rule-based simplification fails to apply directly on the complex Angr generated expressions because of the deeply nested conditionals.
Hence, the algebraic simplification algorithm is essential to simplify the equations by cleaning up the implementation-specific operations.


\eqref{eq:tmon_eq_t}--\eqref{eq:tmon_eq_y3} show the equations generated by REMaQE for the outputs of \texttt{controller\_handler}:

\begin{subequations}
\begin{equation}
t = x_{8} - \frac{x_{1} x_{3}}{x_{4}} \left(x_{0} - 2 x_{6} + x_{7}\right) 
    - x_{1} x_{2} x_{4} \left(x_{0} - x_{5}\right) - x_{1} \left(x_{0} - x_{6}\right) 
\label{eq:tmon_eq_t}
\end{equation}
\begin{equation}
y_0 = \operatorname{round}{\left(\left|y_2\right| \right)}
\label{eq:tmon_eq_y0}
\end{equation}
\begin{equation}
y_1 = \left|{y_2}\right|
\label{eq:tmon_eq_y1}
\end{equation}
\begin{equation}
y_2 = y_5 = \begin{cases} t & \text{for}\: x_5 - x_0 \leq k_2 \:\text{and}\: k_{4} \geq t \:\text{and}\: k_{3} \leq t \\k_{3} & \text{for}\: x_5 - x_0 \leq k_2 \:\text{and}\: k_{3} > t \\k_{4} & \text{for}\: x_5 - x_0 \leq k_2 \:\text{and}\: k_{3} \leq t \:\text{and}\: k_{4} < t \\0 & \text{otherwise} \end{cases}
\label{eq:tmon_eq_y2}
\end{equation}
\begin{equation}
y_3 = y_4 = \begin{cases} x_{0} & \text{for}\: x_5 - x_0 \leq k_2 \\0 & \text{otherwise} \end{cases}
\label{eq:tmon_eq_y3}
\end{equation}
\end{subequations}
where $x_i$ for $i=0, \cdots, 8$ are the inputs, $y_i$ for $i=0,\cdots,5$ are the outputs, $k_i$ for $i=2,3,4$ are constants, and $t$ is a term introduced to simplify the presentation of equations.
The following changes are performed manually to the equations generated by REMaQE:
the term $t$ is introduced and replaces the long shared expression in the outputs;
the expression for $y_2$ is replaced in the $y_0$ and $y_1$ equations;
equations for the repeated outputs $y_2$, $y_5$, and $y_3$, $y_4$ are displayed together.

\subsection{Limitations} \label{sec:assumptions}

REMaQE does not handle data type casting or precision conversion that are intended to perform operations in the original equations (e.g., floor$(x)$ implemented using \texttt{float} to \texttt{int} cast).
Instead, we assume that data type and precision conversions are a by-product of the implementation, and choose to not represent them in the generated equations to avoid clutter.
Similarly, comparison operations that rely on bit manipulation (e.g., $x < 0$ implemented using sign-bit check) are not simplified.
These operations are still supported when performed using the proper instructions or library calls.
Advanced forms of control-flow like function pointers, recursion, or obfuscated control-flow will run into the path explosion problem of symbolic execution, which will impact the ability to analyze such functions.
REMaQE generates equations represented with simple operations on scalar values. 
So, advanced math operations such as vector dot-product or matrix multiplication are unrolled and represented as a long sequence of individual multiply and add operations.
This is less than ideal when the goal is to recover human-friendly equations.
Algebraic simplification cannot handle such subjective modifications to the equations to further simplify them, as also seen in \secref{simplify_eq} (variable $t$ was separated manually).

Despite these limitations, REMaQE is applicable to many real-world reverse engineering scenarios.
Supporting data type conversions, bit-manipulation, vector and matrix operations, will expand the scope of REMaQE to many more applications, and these are targeted for future work.

\section{Evaluation of REMaQE} \label{sec:evaluation}

We focus our evaluation on the widely-used 32-bit ARM architecture with hardware floating point support (ARM32-HF).
It is straightforward to extend REMaQE to other architectures supported by symbolic execution frameworks.
We evaluate REMaQE based on correctness defined in \secref{correctness}, human-friendliness of reverse-engineered equations defined in \secref{friendliness}, and  execution time for reverse engineering.
Dataset generation is described in \secref{dataset}.

\subsection{Correctness} \label{sec:correctness}

This section discusses the guarantees of correctness that the framework provides, along with potential sources of inaccuracy.
A math equation recovered by REMaQE is  ``correct" if it is mathematically equivalent to the original equation.
Equivalence checking is hard especially when dealing with floating point numbers. Two equations $f$ and $g$ are mathematically equivalent if $f(x) = g(x) \:\forall x \in \mathbb{R}^d$, where $d$ is the number of inputs.
As floating point operations are approximate, this strict equality check must be loosened: $\left|\frac{f(x)-g(x)}{f(x)}\right| < \epsilon \:\forall x \in \mathbb{R}^d$, where $\epsilon$ is a tolerance to compare the equation outputs.
The following sources of inaccuracy arise when implementing equations with floating point variables:

\begin{enumerate}
    \item The constants in the original equation are stored in a fixed precision in the binary, hence they can only be recovered to that approximation.
    \item The constants can be modified when the recovered equation is simplified, hence they may not match the original equation constants.
    \item The order of operations can differ between the original equation and recovered equation which impacts floating point computations.
\end{enumerate}
Apart from these inaccuracies, the REMaQE pipeline is deterministic.
Angr's symbolic execution, the algebraic simplification algorithm, and Sympy's rule-based modifications preserve and represent the semantic structure of operations present in the binary.
This ensures REMaQE always recovers a ``correct'' equation, even if it is obscured.
Equivalence check of recovered equations with the original equations is performed at three levels:
\begin{enumerate}
    \item {\bf Structural match} is when they have the same order of operations and variables; This is checked by comparing Sympy's internal tree representation of both equations.
    \item {\bf Semantic match} is when they are equivalent, but do not have the same structure; This is checked by taking the difference of equations and seeing if it simplifies to $0$.
    \item {\bf Evaluated match} is when the functions are evaluated for a range of inputs and the outputs are compared with tolerance $\epsilon = 10^{-5}$.
\end{enumerate}
A structural match implies semantic match, and a semantic match implies evaluated match.
Hence, the equivalence check can stop at the first level that matches.

\subsection{Human Friendliness} \label{sec:friendliness}

A recovered equation may be considered human-friendly if its complexity is similar to the original equation, based on its appearance.
Evaluating the complexity of a math equation is subjective and cannot be precisely measured.
So determining whether an equation is fully simplified is not feasible.
A heuristic measure of complexity is defined by the number of operations in the math equation.
The \texttt{sympy.count\_ops} function measures the number of operations in an equation.
If the reverse engineered equation has a similar number of operations as the original one, we consider them to have comparable complexity.
We quantify human-friendliness as the ratio of the number of operations, with a ratio closer to 1 indicating higher human-friendliness.

\subsection{Dataset Generation} \label{sec:dataset}

Each generated equation is implemented as a C function and a Simulink~\cite{simulink} model.
Simulink is popular for modelling controllers, and it provides features to compile the models into binaries for embedded targets.
The C function is compiled for ARM32-HF target using the GCC compiler (\texttt{arm-linux-gnueabhihf-gcc}). 
The Simulink model is compiled for the same target using Simulink's code generation feature.
Four optimization levels from ``-O0'' to ``-O3'' are used during compilation to obtain a variety of implementations.

To allow direct conversion to Simulink models, which are computational graphs, the math equations are generated as directed-acyclic graphs (DAG), where each node is a parameter or a math operation.
The input nodes are initialized first. Then, math operation nodes are picked from a pool of operations, with repeats allowed.
Edges are randomly added between the nodes to connect the entire graph, while ensuring that  the graph remains a DAG (i.e., no cycles).
Finally, nodes without outgoing edges are connected by inserting add or multiply operation nodes to get to one output.
This final step prevents ``dead'' operations (wasted computation not reaching any output).

REMaQE contrasts with approaches in \cite{Weideman21PERFUME,Lample2020Deep} to randomly generate math equations which generate expression trees instead of DAGs. Generating expression trees helps control the complexity of the random equation.
This is important for the two papers as their analysis uses machine learning and can handle only a finite amount of tokens. REMaQE simplification stage is rule-based and does not have this limitation. 
In our assessment, DAGs offer two advantages:
(1) they can be converted into Simulink models and
(2) they generate complex equations in terms of number of operations.

\tabref{generation_params} shows the parameters available for random DAG generation and compilation.
All combinations of parameters are used so that the dataset has a wide variety of equations for evaluation.
The number of nodes increases to prevent cycles during random connection, so selecting number of nodes up to 15 is enough to create DAGs with up to 25 nodes and equations with 100s of operations,
The trigonometric and exponential operations generate binaries with calls to library functions, which represents real-world implementations.
Conditional operations like saturation, signum, absolute value, and deadzone produce binaries with data-dependent branches and conditional execution that stresses different stages of the REMaQE pipeline.

Each generated equation is simplified first and discarded if the simplification either reduces to a constant, contains complex or infinite values, or cancels out one of the inputs.
Each model in the dataset has a generated equation, a simplified ground truth equation, and binaries compiled using C and Simulink implementations with 4 optimization levels.
The final dataset contains 3,137 ground truth equations and 25,096 compiled binaries for analysis.


\begin{table}[ht]
    \centering
    \caption{Parameters for random DAG generation and binary compilation.}
    \label{tab:generation_params}
    \begin{tabular}{|c|l|}
         \hline
         \textbf{Parameter}& \textbf{DAG generation options} \\ \hline
         Type of operations & Arithmetic, Trigonometric + Exponential, Conditional\\ \hline
         Number of inputs & 1 to 2 \\ \hline
         Number of nodes & 5 to 15 \\ \hline
         & \textbf{Compilation options}\\ \hline
         Implementation & C, Simulink \\ \hline
         Optimization level & -O0, -O1, -O2, -O3 \\ \hline
    \end{tabular}
\end{table}

\section{Results} \label{sec:results}

\begin{table}[bpt]
\small
    \caption{Examples to demonstrate the different types of equivalence checks. Even in instances such as the one shown in the ``Evaluated'' row, REMaQE recovers the correct representation of  equations implemented in the binary (which may differ from ``Ground Truth'') since the compilation tool chain adds these checks.  
    }
\label{tab:equivalence}
\centering
\begin{tabular}{|c|c|c|}
\hline
{\bf Match type} & {\bf Ground Truth} & {\bf Equation Recovered by REMaQE}  \\
\hline
Structural 
& $(-k_1 + x_0\text{sin}(k_0x_1)^2 - x_1)/k_2$
& $(-k_1 + x_0\text{sin}(k_0x_1)^2 - x_1)/k_2$ \\
\hline
Semantic 
& $(-k_0 + k_1 + k_2x_0 - x_1 -\text{atan}(x_0))\text{exp}(-x_0)$
& \(\frac{-\text{atan}(x_0) + x_0 k_2 + k_1 - x_1 - k_0}{\text{exp}(x_0)}\) \\
\hline
Evaluated
& \(\text{acos}(k_1x_0) - \frac{x_1\text{sin}(x_0)}{k_0}\)
& 
\(\begin{cases} 
    \text{acos}(k_1x_0) - x_1\text{sin}(x_0)/k_0 & \\
    \quad\text{for}\: (k_1x_0 >= -1.0) 
 \:\text{and}\: (k_1x_0 <= 1.0) &\\
    3.141593 - x_1\text{sin}(x_1)/k_0 & \\
    \quad \text{for}\: k_1x_1 <= 1.0 & \\
    -x_1\text{sin}(x_0)/k_0 & \\\quad \text{otherwise} &\\ \end{cases}\) 

\\
\hline
\end{tabular}
\end{table}

REMaQE successfully recovers the correct equation for all 25,096 binaries.
This strongly shows the correctness of REMaQE's reverse engineering pipeline.
In the equivalence check, \emph{structural match} is obtained for 12,032 binaries (47.94\%), \emph{semantic match} is obtained for 12,525 binaries (49.91\%), and \emph{evaluated match} is obtained for 533 binaries (2.12\%).
A tolerance of $\epsilon = 10^{-5}$ is used for the \emph{evaluated match}.
The 6 remaining binaries violate the tolerance threshold during evaluation, hence the equivalence check fails for them.
However, manual verification (see \secref{manual_verif}) reveals that REMaQE has recovered the correct equation for these 6 binaries as well.
\figref{maxerr_eval} is a histogram of binaries versus maximum error during \emph{evaluated match}, for the 539 binaries (533 matched plus 6 manually verified).
Most binaries have a maximum error $\leq 10^{-8}$, indicating the equations are correct with confidence.
Very few have higher errors and even fewer breach the tolerance threshold due to floating point approximations.

\begin{minipage}[t]{0.45\linewidth}
\centering
    \includegraphics[width=\textwidth]{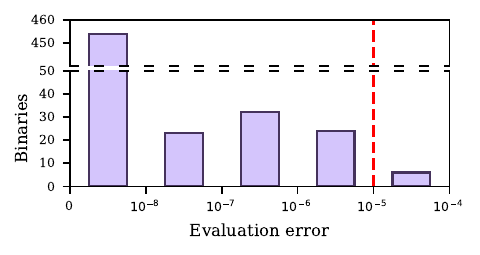}
    \captionof{figure}{Histogram of number of binaries vs. max error during \emph{evaluation match}. The tolerance $\epsilon = 10^{-5}$ is marked.}
    \label{fig:maxerr_eval}
\end{minipage}
\hfill
\begin{minipage}[t]{0.49\linewidth}
\centering
    \includegraphics[width=\textwidth]{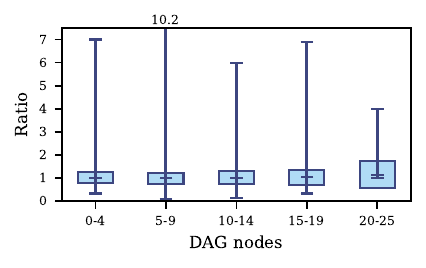}
    \captionof{figure}{Average ratio of equation complexity for each parameter binned with the number of nodes.
    The box indicates $\mu \pm \sigma$ ($\mu$ is mean, $\sigma$ is standard deviation), while the line and whiskers indicate range.}
    \label{fig:complexity_nodes}
\end{minipage}
\vspace{1em}

\tabref{equivalence} shows examples from the dataset for each of the three equivalence check levels.
The recovered equation for the \emph{evaluated match} case is much more complex than the ground truth, with extra conditional operations, because Simulink adds bounds checks for the $\text{acos}(\cdot )$ function.
REMaQE reverse engineers the implementation in the binary, hence semantic modifications performed during compilation, such as the extra bounds checks introduced by Simulink in this case, are also recovered.
The recovered equations help debug the implementation of mathematical algorithms to ensure that the compilation pipeline has not made any unexpected changes.
Evaluating both equations shows that they generate the same outputs (or close, up to the tolerance). 


\subsection{Manual Verification} \label{sec:manual_verif}

The goal of REMaQE is to reverse engineer the binary implementation and present it as math equations.
When the three equivalence checks fail to match the recovered and ground truth equations, this is not sufficient to conclude that the recovered equation is wrong.
Manual inspection is required to determine whether the recovered equation is a faithful recreation of the implementation of the ground truth equation, or if it is indeed mismatched.
The following is one of the 6 manually verified cases to demonstrate that even though equivalence checks have failed, REMaQE has recovered correct equations.
\eqref{eq:fail1_gt} shows the ground truth and \eqref{eq:fail1_rev} shows the recovered equation:
\begin{subequations}
\begin{equation}
    y = \begin{cases}
            x - k_{0} - k_{1} - \frac{1}{k_{0}^{2}} - \frac{x}{k_{0}} + \frac{1}{k_{0}} & \text{for}\: x > 0 \\
            x - k_{0} - k_{1} - \frac{1}{k_{0}^{2}} + \frac{x}{k_{0}} + \frac{1}{k_{0}} & \text{for}\: x < 0 \\
            x - k_{0} - k_{1} & \text{otherwise}
    \end{cases}
\label{eq:fail1_gt}
\end{equation}
\begin{equation}
    y = \begin{cases}
            - \fp{k_0} - \fp{k_1} & \text{for}\: x = 0 \\
            x - \fp{k_0} - \fp{k_1} - \fp{k_2} \left(x - \fp{k_0}+ \fp{k_2} + 1\right) + 1 & \text{for}\: x < 0 \\
            x - \fp{k_0} - \fp{k_1} - 1 + \frac{\fp{k_0} - x + 1 - \frac{1}{\fp{k_0}}}{\fp{k_0}} & \text{for}\: x > 0
    \end{cases}
\label{eq:fail1_rev}
\end{equation}
\end{subequations}
where $x$ is the input, $y$ is the output, $k_i$ are the constants in the ground truth equation, and $\fp{k_i}$ are the approximated constants recovered from the binary.
$k_0=-1.51$, $k_1=-1.58$, $\fp{k_0}=-1.5099999904632568$, $\fp{k_1}=-1.5800000429153442$, and $\fp{k_2}=-0.6622516512870789$.
Notice that \eqref{eq:fail1_rev} has an extra constant $\fp{k_2} \approx \frac{1}{k_0}$, which means the compiler has chosen to store the inverse of $k_0$ as a separate constant.
Inaccuracies due to the floating point approximation and different representation of constants caused \emph{evaluated match} to find inputs that violate the tolerance level.
Manual inspection shows that REMaQE has recovered a semantically matched equation.


\subsection{Human Friendliness}

\figref{complexity_nodes} demonstrates the human-friendliness of reverse engineered outputs of REMaQE.
The average ratio of number of operations in the recovered equation over the simplified ground truth equation is shown as a histogram binned with respect to number of nodes in the DAG.
Standard deviation and range of each bin is displayed as the box and whiskers plot.
The average ratio is close to 1 across the range of equation complexity, with a standard deviation of 0.26; For larger number of nodes, the mean increases to 1.15 and standard deviation increases to 0.57.
REMaQE performs well on the human-friendliness metric for a variety of equations. 
Instances with higher ratios indicate outliers where REMaQE recovers a more complex equation since algebraic simplification cannot shrink the recovered equations beyond a point.
Instances with ratios lower than 1 indicate cases where the recovered equations are simpler than ground truth since C and Simulink compilers optimized the expressions in the implementation. 

\subsection{Execution Time}

\begin{figure*}[!ht]
    \centering
    \includegraphics[width=\linewidth]{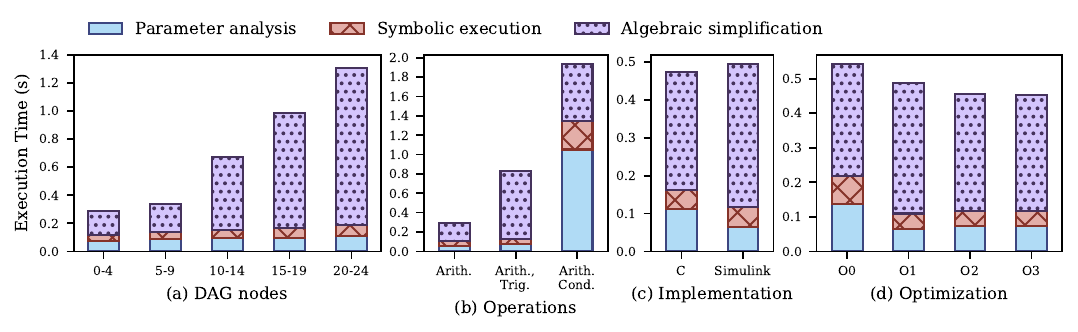}
    \caption{Average execution time (seconds) per stage for the different dataset generation and compilation parameters: (a) number of DAG nodes, (b) type of operations, (c) implementation, and (d) optimization level.}
    \label{fig:runtime}
\end{figure*}

Execution time of each of REMaQE's three stages is measured for the generated dataset on an Intel i7-6700 CPU.
Only one instance of REMaQE is run on a single core to collect the timing measurements.
On average, the full reverse engineering takes 0.48 seconds. 71\% time is consumed by algebraic simplification, 18\% by parameter analysis, and 11\% by symbolic execution.


\figref{runtime} displays the execution time for different dataset generation parameters.
Across the different kinds of binaries, the total execution time remains under 2 seconds.

\noindent\textbf{(a) DAG nodes:}
Execution time for all three stages increases almost linearly with number of nodes in the generated DAG.
This is because code size (i.e. number of instructions) is proportional to DAG nodes.
Execution time ranges from 0.3 seconds to 1.3 seconds across DAG nodes.

\noindent\textbf{(b) Type of operations:}
We see a significant increase for the more complex operations.
Trigonometric and exponential operations cause a larger number of function calls to libraries, taking longer to reverse engineer.
Conditional operations add data-dependent branches that lead to multiple exploration paths during parameter analysis and symbolic execution.
Each branch execution adds conditions to the final equation that require additional handling during algebraic simplification.

\noindent\textbf{(c) Implementation:}
Parameter analysis and symbolic execution are slightly faster for Simulink compared to C, whereas algebraic simplification is slightly slower.
This is because Simulink may add extra bounds checks for some operations that take longer to process during algebraic simplification.

\noindent\textbf{(d) Optimization level:}
For higher optimization levels, parameter analysis and symbolic execution time decreases, as the code size decreases.

Overall, REMaQE reverse engineers a wide range of equations in real time.
It is hence suitable as an interactive tool that can integrate into a GUI-based reverse engineering workflow and deliver recovered equations at latencies comparable to human click speeds.


\section{Case Study: Reverse Engineering of ArduPilot} \label{sec:ardupilot}

We use REMaQE to reverse engineer two functions in the ArduPilot auto-pilot firmware for unmanned aerial vehicles \cite{ArduPilot}.
The firmware binary\footnote{\url{https://firmware.ardupilot.org/Copter/stable-4.3.4/bbbmini/arducopter}} is compiled for ARM32-HF target for the BeagleBone Black embedded platform.
The binary is not stripped in this case, so function names and global symbol names are available. 
\figref{ardupilot_cpp} shows the C++ source for \texttt{PosVelEKF::predict} which implements an extended Kalman filter (EKF) for position-velocity estimation and \texttt{calc\_lowpass\_alpha\_dt}, a low-pass filter.
We reverse engineer these functions using REMaQE.
The source code is provided only for discussion and is not utilized during reverse engineering.

\begin{figure}[tpb]
\begin{lstlisting}[numbers=none,language=C++]
void PosVelEKF::predict(float dt, float dVel, float dVelNoise)
{
    // Newly predicted state and covariance matrix at next time step
    float newState[2];
    float newCov[3];
    // We assume the following state model for this problem
    newState[0] = dt*_state[1] + _state[0];
    newState[1] = dVel + _state[1];
    /* ... */
    newCov[0] = dt*_cov[1] + dt*(dt*_cov[2] + _cov[1]) + _cov[0];
    newCov[1] = dt*_cov[2] + _cov[1];
    newCov[2] = ((dVelNoise)*(dVelNoise)) + _cov[2];
    // store the predicted matrices
    memcpy(_state,newState,sizeof(_state));
    memcpy(_cov,newCov,sizeof(_cov));
}

float calc_lowpass_alpha_dt(float dt, float cutoff_freq)
{
    if (is_negative(dt) || is_negative(cutoff_freq)) {
        INTERNAL_ERROR(AP_InternalError::error_t::invalid_arg_or_result);
        return 1.0;
    }
    if (is_zero(cutoff_freq)) return 1.0;
    if (is_zero(dt)) return 0.0;
    float rc = 1.0f / (M_2PI * cutoff_freq);
    return dt / (dt + rc);
}
\end{lstlisting}
\caption{C++ source of \texttt{PosVelEKF::predict} and \texttt{calc\_lowpass\_alpha\_dt}.}
\label{fig:ardupilot_cpp}
\end{figure}

\tabref{ardupilot_params} shows the parameter metadata extracted for \texttt{PosVelEKF::predict}.
REMaQE correctly identifies the 3 pass-by-value arguments ($x_0$ to $x_2$) and 5 class members ($x_5$ to $x_9$) as inputs of the function.
The missing indices are false positive inputs identified by REMaQE. They do not show up in the output equations, and they are not displayed in the table.
The class members are also identified as the 5 outputs ($y_0$ to $y_4$).
The class pointer is correctly identified as the base for the class members and represented as \texttt{ptr0}.
REMaQE has handled the \texttt{memcpy} call and determined outputs even though the function returns \texttt{void}.

\begin{table}[htpb]
\caption{Parameters of \texttt{PosVelEKF::predict}.}
\label{tab:ardupilot_params}

\begin{minipage}{0.45\linewidth}
\centering
\begin{tabular}{|l|l|l|r|}
\hline
 \textbf{Input}   & \textbf{Kind}    & \textbf{Location}        &   \textbf{Size} \\
\hline
 $x_0$     & register& \texttt{s0}&     32 \\
 $x_1$     & register& \texttt{s1}&     32 \\
 $x_2$     & register& \texttt{s2}&     32 \\
 $x_5$     & pointer & \texttt{ptr0[0x0] } &     32 \\
 $x_6$     & pointer & \texttt{ptr0[0x4] } &     32 \\
 $x_7$     & pointer & \texttt{ptr0[0x8] } &     32 \\
 $x_8$     & pointer & \texttt{ptr0[0xc] } &     32 \\
 $x_9$     & pointer & \texttt{ptr0[0x10]} &     32 \\
\hline
\end{tabular}    
\end{minipage}
\begin{minipage}{0.45\linewidth}
\centering
\begin{tabular}{|l|l|l|r|}
\hline
 \textbf{Output}   & \textbf{Kind}    & \textbf{Location}  &   \textbf{Size} \\
\hline
$y_0$     & pointer & \texttt{ptr0[0x0] } &     32 \\ 
$y_1$     & pointer & \texttt{ptr0[0x4] } &     32 \\ 
$y_2$     & pointer & \texttt{ptr0[0x8] } &     32 \\ 
$y_3$     & pointer & \texttt{ptr0[0xc] } &     32 \\ 
$y_4$     & pointer & \texttt{ptr0[0x10]} &     32 \\
\hline
 \textbf{Pointer}   & \textbf{Kind}   & \textbf{Location}   &   \textbf{Size} \\
\hline
    \texttt{ptr0} & register & \texttt{r0}        &     32 \\
\hline
\end{tabular}

\end{minipage}
\end{table}

\eqref{eq:ardupilot_out1}--\eqref{eq:ardupilot_out5} show the recovered output equations:
\begin{subequations}
\begin{equation}
    y_0 = x_{0} x_{6} + x_{5} \label{eq:ardupilot_out1}
\end{equation}
\begin{equation}
    y_1 = x_{1} + x_{6}
\end{equation}
\begin{equation}
    y_2 = x_{0} x_{8} + x_{0} \left(x_{0} x_{9} + x_{8}\right) + x_{7}
\end{equation}
\begin{equation}
    y_3 = x_{0} x_{9} + x_{8}
\end{equation}
\begin{equation}
    y_4 = x_{2}^{2} + x_{9} \label{eq:ardupilot_out5}
\end{equation}
\end{subequations}
where $x_i$ for $i=0,1,2,5,\cdots,9$ are inputs and $y_i$ for $i=0,\cdots,4$ are outputs.
REMaQE  reverse engineers the math equations implemented in the function. While these equations by themselves are not sufficient to determine that this function implements an EKF, the user can combine them with context from other sources to understand aspects like the state and covariance updates.

\tabref{ardupilot_lowpass_params} shows the parameter metadata extracted for \texttt{calc\_lowpass\_alpha\_dt}.
The error function is ignored during symbolic execution so REMaQE captures the error cases without terminating.
REMaQE identifies the pass-by-value arguments and return output. The recovered constant values are helpful: $k_0\approx2\pi$, also $k_1\approx -2^{-23}$ and $k_2\approx 2^{-23}$, the 32-bit floating point machine epsilons.
\begin{table}[tb]
\caption{Parameters of \texttt{calc\_lowpass\_alpha\_dt}.}
\label{tab:ardupilot_lowpass_params}

\begin{minipage}{0.43\linewidth}
\centering
\begin{tabular}{|l|l|l|r|}
\hline
 \textbf{Input}   & \textbf{Kind}   & \textbf{Location}   &   \textbf{Size} \\
\hline
 $x_0$ & register & \texttt{s0}  &     32 \\
 $x_1$ & register & \texttt{s1}  &     32 \\
\hline
 \textbf{Output}   & \textbf{Kind}   & \textbf{Location}   &   \textbf{Size} \\
\hline
 $y_0$ & register & \texttt{s0}   &     32 \\
\hline
\end{tabular}
\end{minipage}
\begin{minipage}{0.54\linewidth}
\centering
\begin{tabular}{|l|l|l|r|r|}
\hline
 \textbf{Constant}   & \textbf{Kind}   & \textbf{Location}   &   \textbf{Size} &        \textbf{Value} \\
\hline
 $k_0$ & global & \texttt{0x473f88}   &     64 &  6.28319     \\
 $k_1$ & global & \texttt{0x473f90}   &     32 & -1.19209e-07 \\
 $k_2$ & global & \texttt{0x473f94}   &     32 &  1.19209e-07 \\
\hline
\end{tabular}
\end{minipage}

\end{table}

\eqref{eq:lowpass} shows the output equation:
\begin{equation}
y_0 = \begin{cases} 
1 & \text{for}\: k_{1} \leq x_{0} \:\text{and}\: k_{1} \leq x_{1} 
    \:\text{and}\: k_{2} > \left|{x_{1}}\right| \\
0 & \text{for}\: k_{2} \geq \left|{x_{0}}\right| \:\text{and}\: k_{1} \leq x_{0} 
    \:\text{and}\: k_{1} \leq x_{1}  \:\text{and}\: k_{2} \leq \left|{x_{1}}\right| \\
\frac{k_{0} x_{0} x_{1}}{k_{0} x_{0} x_{1} + 1} & \text{for}\: k_{1} \leq x_{0} \:\text{and}\: k_{1} \leq x_{1} 
    \:\text{and}\: k_{2} \leq \left|{x_{1}}\right| \:\text{and}\: k_{2} < \left|{x_{0}}\right| \\
1 & \text{otherwise} 
\end{cases}
\label{eq:lowpass}
\end{equation}
where $y_0$ is the output, $x_0$, $x_1$ are the inputs, and $k_0$, $k_1$, $k_2$ are constants.
The expression for low-pass filter is shown in one of the cases.
The conditions in  case statements indicate that \texttt{is\_zero()} and \texttt{is\_negative()} in the original code are implemented as comparisons with $k_1$ and $k_2$, the machine epsilon, instead of relying on equality or sign-bit check. Using REMaQE, we verify that the implementation is robust to floating point approximations. 
This case study demonstrates that REMaQE can be used in real-world analysis of UAVs, and debugging of implemented algorithms.

\section{Case Study: Reverse Engineering of the OpenPLC PID controller} \label{sec:openplc}

We use REMaQE to reverse engineer a PID controller implemented using the OpenPLC~\cite{OpenPLC} platform in Structured Text (ST) language.
OpenPLC converts the ST implementation to C code, which is compiled to an object file for the ARM32-HF target using the GCC compiler.
This object file can be executed using the OpenPLC runtime on embedded devices that OpenPLC supports, such as the Arduino and Raspberry Pi.
\figref{pid_st} shows the ST source code of the PID controller. The function blocks in this ST are based on the ``PID'', ``INTEGRAL'', and ``DERIVATIVE''
examples provided in the IEC~61131-3 standard~\cite{IEC61131}. The source code is provided here only for discussion and is not used during reverse engineering.

\begin{figure}[htpb]
\begin{subfigure}[t]{0.45\textwidth}
    \begin{lstlisting}[language=ST]
FUNCTION_BLOCK MYDERIVATIVE
  VAR_INPUT xin:REAL; cycle:REAL; END_VAR
  VAR_OUTPUT xout:REAL; END_VAR
  VAR x1, x2, x3:REAL; END_VAR
  
  xout := (3.0 * (xin - x3) + x1 - x2) 
          / (10.0 * cycle);
  x3 := x2; x2 := x1; x1 := xin;
END_FUNCTION_BLOCK

FUNCTION_BLOCK MYINTEGRAL
  VAR_INPUT xin:REAL; cycle:REAL; END_VAR
  VAR_OUTPUT xout:REAL; END_VAR
  
  xout := xout + xin * cycle;
END_FUNCTION_BLOCK
    \end{lstlisting}
\end{subfigure}
\hfill
\begin{subfigure}[t]{0.48\textwidth}
    \begin{lstlisting}[language=ST,firstnumber=17]
FUNCTION_BLOCK MYPID
  VAR_INPUT pv:REAL; setp:REAL; cycle:REAL; END_VAR
  VAR_OUTPUT xout:REAL; END_VAR
  VAR
    Kp:REAL; Tr:REAL; Td:REAL; error:REAL;
    iterm:MYINTERGRAL; dterm:MYDERIVATIVE;
  END_VAR

  Kp := 1.54; Tr := 2.33; Td := 0.07;
  error := pv - setp;
  iterm (xin := error, cycle := cycle);
  dterm (xin := error, cycle := cycle);
  xout := Kp * (error + iterm.xout/Tr 
                + dterm.xout*Td);
END_FUNCTION_BLOCK
    \end{lstlisting}
\end{subfigure}
    
    \caption{Structured Text implementation of a PID controller for OpenPLC.}
    \label{fig:pid_st}
\end{figure}

The ``MYPID'' block is implemented as the \texttt{MYPID\_body\_\_} function in the object file.
The ``MYINTEGRAL'' and ``MYDERIVATIVE'' blocks also get similar functions, but REMaQE is run directly on \texttt{MYPID\_body\_\_}
as we intend to inline the integral and derivative calculations in the recovered equations instead of representing them as function calls.
\tabref{pid_st_params} shows the parameters of the PID controller. We see that access to input, output and local variables are pointer based.
We also see the three tuning constants used by the controller, whose usage will be evident from the recovered equations.
The inputs $x_4, x_5, x_6, x_7$ and outputs  $y_4, y_5, y_6, y_7$ are correspondingly accessed from the same location (same pointer offset), indicating that they
may refer to the local variables.

\begin{table}[htpb]
\caption{Parameters of the OpenPLC PID controller.}
\label{tab:pid_st_params}
\begin{minipage}{0.45\linewidth}
\centering
    \begin{tabular}{|l|l|l|r|}
\hline
 \textbf{Input}   & \textbf{Kind}   & \textbf{Location}   &   \textbf{Size} \\
\hline
 $x_0$ & pointer & \texttt{ptr0[0x0] } &     32 \\
 $x_1$ & pointer & \texttt{ptr0[0x8] } &     32 \\
 $x_2$ & pointer & \texttt{ptr0[0x10]} &     32 \\
 $x_3$ & pointer & \texttt{ptr0[0x18]} &     32 \\
 $x_4$ & pointer & \texttt{ptr0[0x60]} &     32 \\
 $x_5$ & pointer & \texttt{ptr0[0x80]} &     32 \\
 $x_6$ & pointer & \texttt{ptr0[0x88]} &     32 \\
 $x_7$ & pointer & \texttt{ptr0[0x90]} &     32 \\
\hline
 \textbf{Pointer}   & \textbf{Kind}   & \textbf{Location}   &   \textbf{Size} \\
\hline
 \texttt{ptr0} & register    & \texttt{r0}         &     32 \\
\hline
\end{tabular}
\end{minipage}
\begin{minipage}{0.53\linewidth}
\centering
\begin{tabular}{|l|l|l|r|r|}
\hline
 \textbf{Input}   & \textbf{Kind}   & \textbf{Location}   &   \textbf{Size} & \\
\hline
 $y_0$ & pointer & \texttt{ptr0[0x20]} &     32 & \\
 $y_4$ & pointer & \texttt{ptr0[0x60]} &     32 & \\
 $y_5$ & pointer & \texttt{ptr0[0x80]} &     32 & \\
 $y_6$ & pointer & \texttt{ptr0[0x88]} &     32 & \\
 $y_7$ & pointer & \texttt{ptr0[0x90]} &     32 & \\
\hline
 \textbf{Input}   & \textbf{Kind}   & \textbf{Location}   &   \textbf{Size} & \textbf{Value} \\
\hline
 $k_0$ & immediate    & \texttt{0x40019b}   &     32 &    1.54 \\
 $k_1$ & immediate    & \texttt{0x4001af}   &     32 &    2.33 \\
 $k_2$ & immediate    & \texttt{0x4001c3}   &     32 &    0.07 \\
\hline
\end{tabular}
\end{minipage}
\end{table}

\eqref{eq:pid_st_out1}--\eqref{eq:pid_st_out5} show the recovered output equations:
\begin{subequations}
\begin{equation}
y_0 = k_0 \left(
    x_0 - x_1
    + \frac{1}{k_1} \left(x_4 + x_3 \left(x_0 - x_1\right)\right)
    + \frac{0.1 k_2}{x_3} \left(x_5 - x_6 + 3.0 \left(x_0 - x_1 - x_7 \right)\right)
\right) \label{eq:pid_st_out1}
\end{equation}
\begin{equation}
y_4 = x_4 + x_3 \left(x_0 - x_1\right)  \label{eq:pid_st_out2}
\end{equation}
\begin{equation}
y_5 = x_0 - x_1
\end{equation}
\begin{equation} 
y_6 = x_5
\end{equation}
\begin{equation}
y_7 = x_6   \label{eq:pid_st_out5}
\end{equation}
\end{subequations}
where $x_i$ for $i=0,\cdots,7$ are inputs, $y_i$ for $i=0,4,\cdots,7$ are outputs, and $k_0, k_1, k_2$ are constants.
The equation for $y_0$ looks like the PID output, which means that $k_0$ must be ``Kp''.
The expression $x_0 - x_1$ is repeated in multiple places, indicating that it is the error term, making $x_0$ as process value ``pv'' and $x_1$ as setpoint ``setp''.
As $x_4$ and $y_4$ are stored in the same location, \eqref{eq:pid_st_out2} represents an increment of $x_4$ with $x_3 (x_0 - x_1)$, so it is likely the integral accumulator.
This means the second term in \eqref{eq:pid_st_out1} is the integral term and the third is the derivative term.
So, $y_5, y_6, y_7$ store previous error values in shifted manner, to be used in the four-point derivative approximation term.
We can also conclude that $k_1$ is ``Tr'', $k_2$ is ``Td'', and $x_3$ is the cycle time input.
These findings match the ST implementation in \figref{pid_st}.
Notice that $x_2$ is identified as an input, but is not used in the equations.
This happens when some locations are accessed for other purposes, but parameter analysis marks them as inputs of the equations.
This case study demonstrates that REMaQE can be used to reverse engineer PLC binary executables, which helps to recover source code of legacy systems, and verify the integrity of
implemented equations to ensure there is no tampering.

\section{Conclusion} \label{sec:conclusion}

The REMaQE framework automatically reverse engineers math equations from binary executables. 
It has three stages: parameter analysis, symbolic execution, and algebraic simplification. 
Parameter analysis allows REMaQE to identify the input, output and constant parameters of the math equation
and whether they are stored in the register, on the stack, in global memory or accessed via pointers.
REMaQE  uses this metadata to reverse engineer a wide variety of implementations, such as C++ classes that use class pointers, or Simulink compiled binaries that use global memory.
This is a significant improvement over state-of-the-art that handles simple functions implemented using registers.
Algebraic simplification allows REMaQE to simplify complex conditional expressions involving floating-point comparisons,
that are generated by symbolic execution even for simple equations.
Existing approaches use machine learning models that have limits on the input expression size. They cannot handle the long, complex expressions generated by conditional computations.

Additionally, we introduce an alternative method for randomly generating diverse equations using directed acyclic graphs. 
REMaQE recovers the correct equations for the entire dataset of 25,096 binaries.
REMaQE takes an average execution time of 0.48 seconds and up to 2 seconds for the complex equations.
Such a small reverse engineering time makes REMaQE effective in an interactive reverse engineering workflow,
and would enable one-click latency when integrated with GUI frameworks.
Comparing the complexity of the recovered equations with the original equations, REMaQE shows an average ratio of 1 through a range of
equation complexities, with a standard deviation of 0.26.  The REMaQE automated tool can significantly reduce human effort in the recovery of control system dynamics from legacy hardware or recovered adversaries’ control computers and can also facilitate defenses during run-time or after an attack to determine if any parameters or control dynamics are modified. This can also reveal  the effects of the modifications on the mathematical computations and therefore system performance. Thus, REMaQE extracts human-friendly equations.

Future work can extend REMaQE to integrate it into  GUI decompilation tools, handle data type conversion and bitwise computation, bundle support for complex control-flow like function pointers or recursion, and enhance reverse engineering capabilities to recognize advanced mathematical structures for improved representation.
Algorithms in embedded systems  can employ advanced mathematical structures such as dynamic loops, summation, integration, differentiation, and high dimensional data such as vectors and matrices. 
Extending simplification to include long chains of scalar computation into high-level operations will further improve REMaQE's human-friendliness.
These refinements and extensions are the proposed future work for REMaQE.

\bibliographystyle{ACM-Reference-Format}
\bibliography{main}


\end{document}